\documentclass{jetpl}
\twocolumn

\rus

\usepackage{caption}
\usepackage{graphicx}
\usepackage{subcaption}
\usepackage{cite}
\usepackage[dvips]{graphicx}
\DeclareGraphicsExtensions{.pdf,.png,.jpg,.eps,.ps}
\RequirePackage{caption}
\DeclareCaptionLabelSeparator{point}{. }
\makeatletter
\newcommand{\keepwithnext}{\@beginparpenalty 10000}
\makeatother

\title{Нерегулярность в спектре массового состава первичных космических лучей при энергии $\mathbf{\sim10}$~ПэВ}
\rtitle{Нерегулярность в спектре массового состава\ldots}
\sodtitle{Нерегулярность в спектре массового состава первичных космических лучей при энергии $\mathbf{\sim10}$~ПэВ}

\author{С.\,Е.\,Пятовский \/\thanks{e-mail: vgsep@ya.ru}}
\rauthor{С.\,Е.\,Пятовский}
\sodauthor{Пятовский}

\address{Физический институт им.~П.~Н.~Лебедева~РАН, 119991 Россия, Москва}

\abstract{Рассмотрена нерегулярность в спектре массового состава первичного космического излучения (ПКИ) при энергии $\sim 10$~ПэВ. Для оценки изменения массового состава ПКИ применен метод рентген-эмульсионных камер (РЭК) и метод гало, основанный на методе РЭК. Изучение изменения массового состава ПКИ выполнено по экспериментально полученным характеристикам стволов широких атмосферных ливней (ШАЛ), где флуктуации данных характеристик минимальны и максимально сохранена информация о первичном взаимодействии ядер ПКИ с атомами атмосферы. Показано, что около $E_0=10$~ПэВ наблюдается локальный максимум доли тяжелых ядер. Данный максимум соотнесен с источниками ПКИ, - звездами переменного типа SR (красные гиганты и сверхгиганты) и WR (Вольфа-Райе).\newline
\textbf{Ключевые слова:} первичные космические лучи, массовый состав, звезды переменного типа}

\PACS{96.40.De, 96.40.Pq, 13.85.Tp, 13.85.-t}

\begin{document}

\maketitle
{\bf 1.~Введение}\newline
Изменение массового состава ПКИ при высоких энергиях остается предметом научных дискуссий. Доля протонов при $E_0=1-100$~ПэВ оценивается от 5\% до 20\%, доля легких ядер \textit{р}+Не, - до 70\% и выше, в зависимости от условий эксперимента и модели реконструкции развития ШАЛ в атмосфере~\cite{1}. Например, доля легких ядер в массовом составе ПКИ в модели локального источника Vela Ерлыкина-Павлюченко~\cite{2} при $E_0=3-5$~ПэВ оценивается $\cong$~88\%, в то время как по данным эксперимента \mbox{KASCADE-Grande}~\cite{3} доля протонов в диапазоне $E_0=1-100$~ПэВ не превышает 10\%. Одновременно, результаты эксперимента ARGO-YBJ~\cite{4} показали, что доля легких ядер начинает уменьшаться уже при $E_0=1$~ПэВ и массовый состав ПКИ существенно утяжеляется.

Для оценки массового состава ПКИ особый интерес представляет изучение характеристик стволов ШАЛ, где данные характеристики максимально чувствительны к массовому составу, а также имеют минимальные флуктуации при развитии ядерно-электромагнитного каскада (ЯЭК) в атмосфере. Стволы ШАЛ изучаются методом РЭК, который в настоящее время не имеет аналогов по пространственному разрешению регистрируемых $\gamma$-квантов. В экспериментах с РЭК при $E_0$ около 10~ПэВ наблюдаются события, лишь недавно получившие свое объяснение. В первую очередь, это семейства $\gamma$-квантов с гало, или просто гало. В научных исследованиях~\cite{5} показано, что гало больших площадей $>500$~мм$^2$ образованы перекрытием подпороговых $\gamma$-квантов. Успешное моделирование всего спектра площадей гало позволило применить разработанный метод гало~\cite{6} для оценки массового состава ПКИ как малозависимый от моделирования прохождения ШАЛ через атмосферу. В настоящее время обработка данных РЭК эксперимента \mbox{ПАМИР} продолжается.

{\bf 2.~Метод РЭК}\newline
Для изучения стволов ШАЛ необходима установка с высоким координатным разрешением порядка десятка микрон и площадью десятки кв.~м. Метод РЭК~\cite{7}, - единственный, позволяющий решить данную задачу. Он является переходным между прямыми и модельно-зависимыми методами изучения событий в ШАЛ, образованных ядрами ПКИ с $E_0 \ge 100$~ТэВ. Результат экспериментов с РЭК, - экспонированные рентгенографические пленки (РГП), на которых после проявки регистрируются области потемнения, т.н. $\gamma$-кванты. Измерение уровней потемнений \textit{D} позволяет делать выводы об энергии электронно-фотонной компоненты (ЭФК) ШАЛ.

\begin{figure}[h]
\centering
\includegraphics[width=\linewidth]{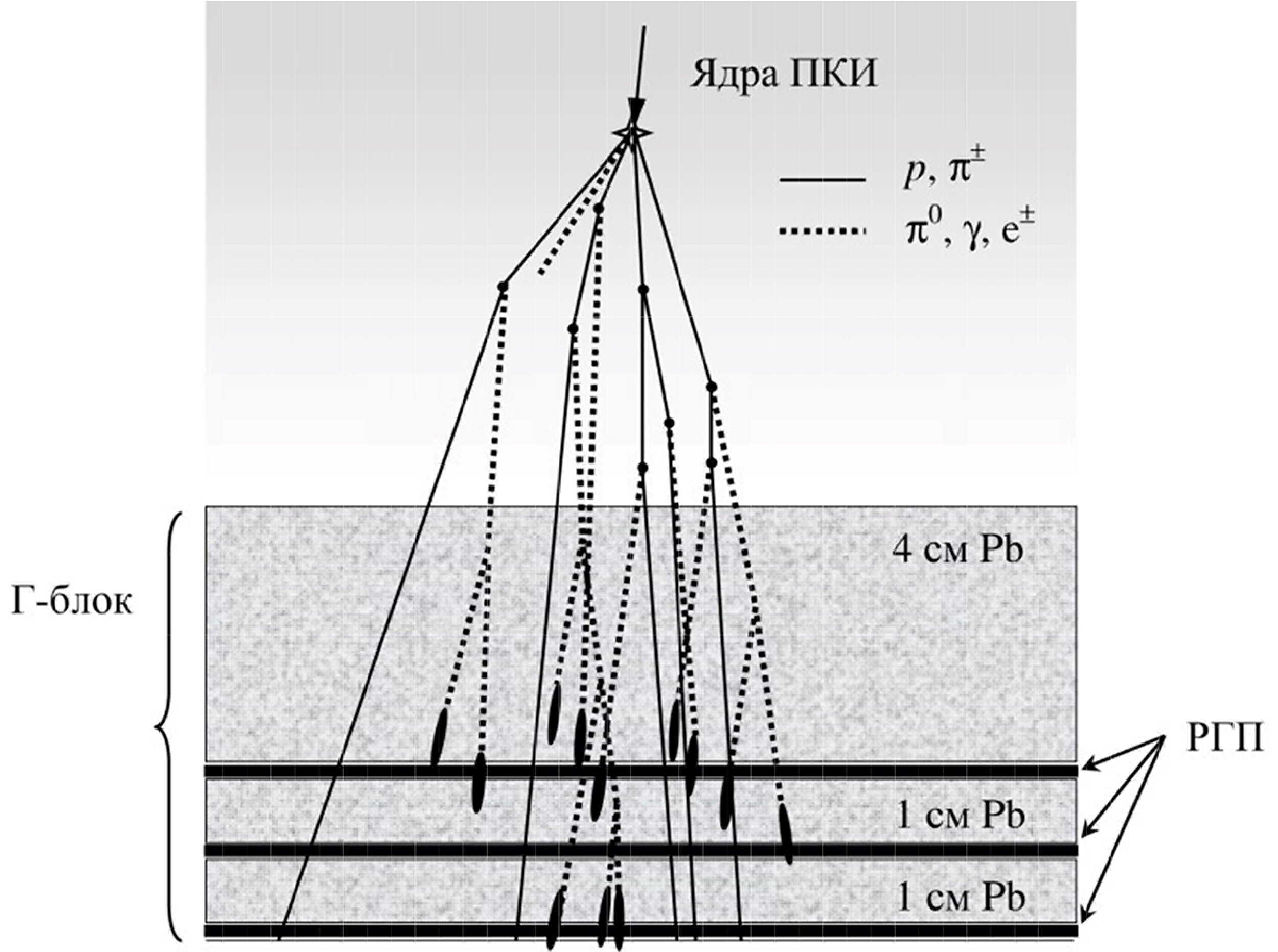}
\caption{Схема Г-блока РЭК эксперимента \mbox{ПАМИР}.}
\label{fig1}
\end{figure}

Схема РЭК эксперимента \mbox{ПАМИР} приведена на рисунке~\ref{fig1}. Цель эксперимента, - получение статистики высокоэнергичных семейств $\gamma$-квантов в стволах ШАЛ и анализ характеристик данных семейств $\gamma$-квантов, образованных ядрами ПКИ с $E_0 \ge 0,1$~ПэВ. Полная экспозиция РЭК эксперимента \mbox{ПАМИР} составила $ST=3000$~м$^2\cdot$год$\cdot$ср на глубине атмосферы $H_\textrm{ПАМИР}=594$~г/см$^2$.

Г-блок РЭК состоит из чередующихся слоев Pb и предназначен для регистрации ЭФК ШАЛ. ЭФК, проходя через Pb, образует вторичные каскады, оставляющие на РГП области потемнений. $\gamma$-кванты фотометрируются, что позволяет оценить энергию $E_\gamma$ частиц ЭФК ШАЛ, вызвавших электромагнитные каскады в Pb.

Исследования показали, что семейства $\gamma$-квантов образованы преимущественно протонами и, в меньшей степени, ядрами Не, что позволяет изучать визуально наблюдаемые характеристики семейств $\gamma$-квантов на РГП как стволы ШАЛ, образованных легкой компонентой ПКИ.

{\bf 3.~Экспериментальные данные РЭК}\newline
Проанализированные экспериментальные данные РЭК содержат характеристики 1294 семейств $\gamma$-квантов и 29112 $\gamma$-квантов, входящих в семейства. В рамках решения поставленных задач изучены такие характеристики семейств $\gamma$-квантов, как зенитный угол $\theta$ прихода ШАЛ, координаты и энергии $E_\gamma$ отдельных $\gamma$-квантов, энергии семейств $\gamma$-квантов $\Sigma E_\gamma$. Минимальная измеренная энергия $\gamma$-квантов составила 4~ТэВ, минимальная $\Sigma E_\gamma$, измеренная по данным $\gamma$-квантам, - 100~ТэВ.

\begin{figure*}[h]
\centering

\begin{subfigure}[t]{0.3\linewidth}
\includegraphics[width=3.8cm]{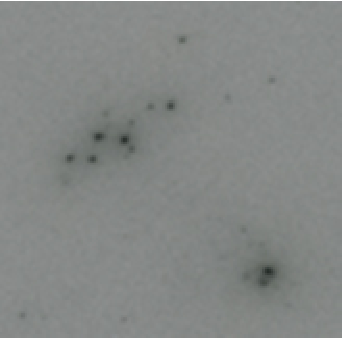}
\centering
\caption{$\Sigma E_\gamma=882$~ТэВ,\newline $S_{D=0,5}=(2,9 + 2,4)$~мм$^2$ (многоцентровое гало, большие $p_\bot$).}
\end{subfigure}
\hspace{.5cm}
\begin{subfigure}[t]{0.3\linewidth}
\includegraphics[width=3.9cm]{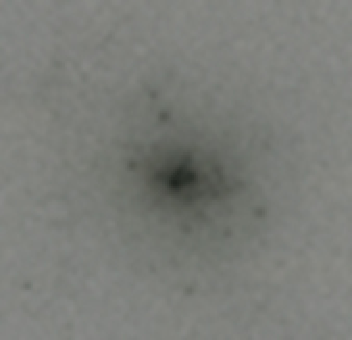}
\centering
\caption{$\Sigma E_\gamma=1084$~ТэВ,\newline $S_{D=0,5}=5,5$~мм$^2$.}
\end{subfigure}
\hspace{.5cm}
\begin{subfigure}[t]{0.3\linewidth}
\includegraphics[width=4cm]{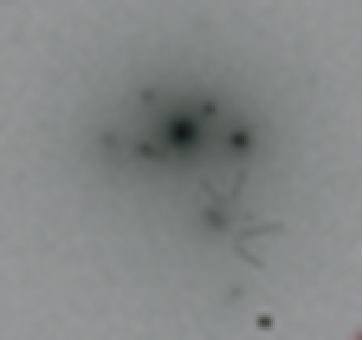}
\centering
\caption{$\Sigma E_\gamma=1688$~ТэВ,\newline $S_{D=0,5}=46,1$~мм$^2$.}
\end{subfigure}
\caption{Гало на РГП, полученные в эксперименте с РЭК.}
\label{fig2}
\end{figure*}

В исследованиях~\cite{1,5} показано, что энергия $E_0$ в несколько ПэВ является порогом образования гало, - больших пятен потемнения на РГП. Примеры событий с гало приведены на рисунке~\ref{fig2}.

Площадь гало может достигать площади 1000 и более кв.~мм. Исследования показали, что максимальная площадь гало на глубине эксперимента РЭК~\mbox{ПАМИР} (594~г/см$^2$ атмосферы) не может превышать 4000~мм$^2$ по причине того, что в данном случае ЯЭК не успевает развиться. Однако такие события чрезвычайно редки, и максимально зарегистрированная площадь гало составила 1450~мм$^2$ с $\Sigma E_\gamma=8260$~ТэВ, что соответствует $E_0>80$~ПэВ.

Природа гало больших площадей объяснена в работах~\cite{8}, где показано, что данные события не относятся к экзотическим и образованы перекрытием функций пространственного распределения (ФПР) подпороговых $\gamma$-квантов. Так как методом РЭК регистрируются преимущественно ШАЛ, образованные ядрами легкой компоненты ПКИ, гало представляют собой рентген-изображения стволов ШАЛ, образованных ядрами \textit{р}+Не ПКИ. Моделирование гало также позволило разработать и апробировать на экспериментальных данных малозависимый от условий моделирования ШАЛ метод гало~\cite{6}, позволяющий оценивать долю \textit{р}+Не в массовом составе ПКИ.

{\bf 4.~Моделирование экспериментальных данных РЭК}\newline
Программный комплекс, разработанный для моделирования экспериментальных данных РЭК, объединяет результаты расчетов по модели МС0-FANSY~\cite{9,10}, верифицированной по данным РЭК эксперимента \mbox{ПАМИР} и актуализированной по экспериментальными данными LHCf, а также результаты расчетов по моделированию ФПР е$^\pm$ и $\gamma$-квантов при прохождении ШАЛ через Г-блок РЭК~\cite{11}.

К параметрам ШАЛ, моделируемым на уровне РЭК, отнесены тип первичного ядра (\textit{р}, He, Li, C, O, Mg, Si, V, Fe), $E_0$, $cos\theta$, энергии и $x,y$-координаты отслеживаемых частиц в плоскости РЭК. Пороговая энергия отслеживаемых в ШАЛ частиц принята равной 100~ГэВ c моделированием генераций более 300 адронов. ФПР получены с учетом эффекта Ландау-Померанчука-Мигдалла, включая осевое приближение для малых $r$ и больших энергий лавинных е$^\pm$.

\begin{figure}[h]
\centering
\includegraphics[width=\linewidth]{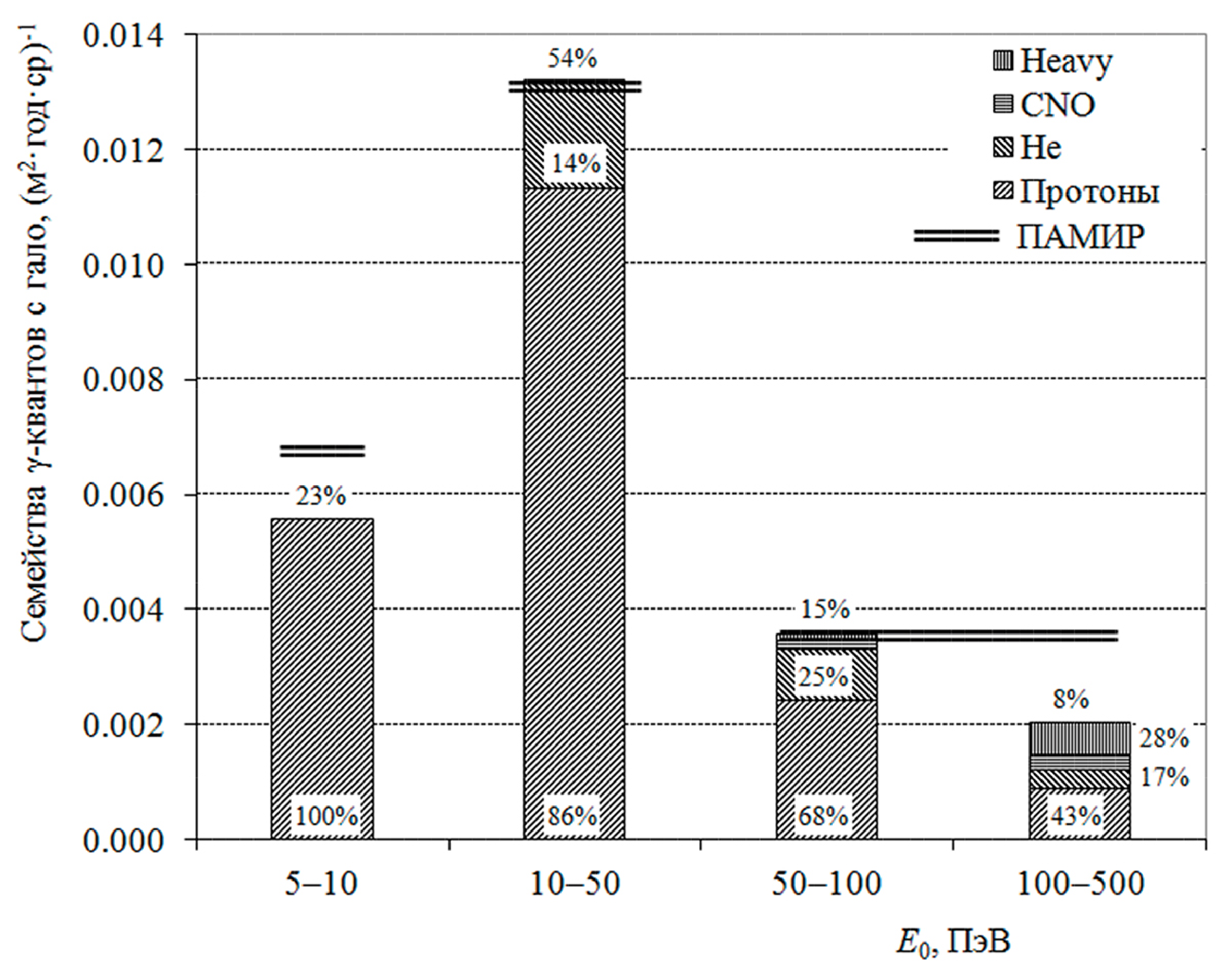}
\caption{Доля гало, образованных различными ядрами ПКИ в сравнении с данными эксперимента \mbox{ПАМИР}.}
\label{fig3}
\end{figure}

На рисунке~\ref{fig3} приведены зависимости потоков событий с гало от $E_0$, полученных в эксперименте \mbox{ПАМИР}. Из рисунка~\ref{fig3} следует хорошее согласие расчетного и экспериментально полученного потоков гало, а также то, что гало образованы преимущественно легкими ядрами массового состава ПКИ. Из рисунка~\ref{fig3} также следует, что пороговая энергия образования гало составляет несколько~ПэВ. Начиная с энергий $E_0\sim100$~ПэВ в образовании гало начинают участвовать все ядра ПКИ, от протонов до ядер Fe. Однако в силу степенного спектра по $E_0$ ПКИ поток событий, регистрируемых в РЭК, при $E_0=100$~ПэВ относительно событий при более низких энергиях незначителен и оценка массового состава ПКИ по событиям с гало относится к $E_0=10$~ПэВ как средневзвешенной по вероятностям образования гало различными ядрами ПКИ.

{\bf 5.~Доля легких ядер при $\mathbf{\textit{E}_0\sim10}$~ПэВ}
\begin{figure}[h]
\centering
\includegraphics[width=\linewidth]{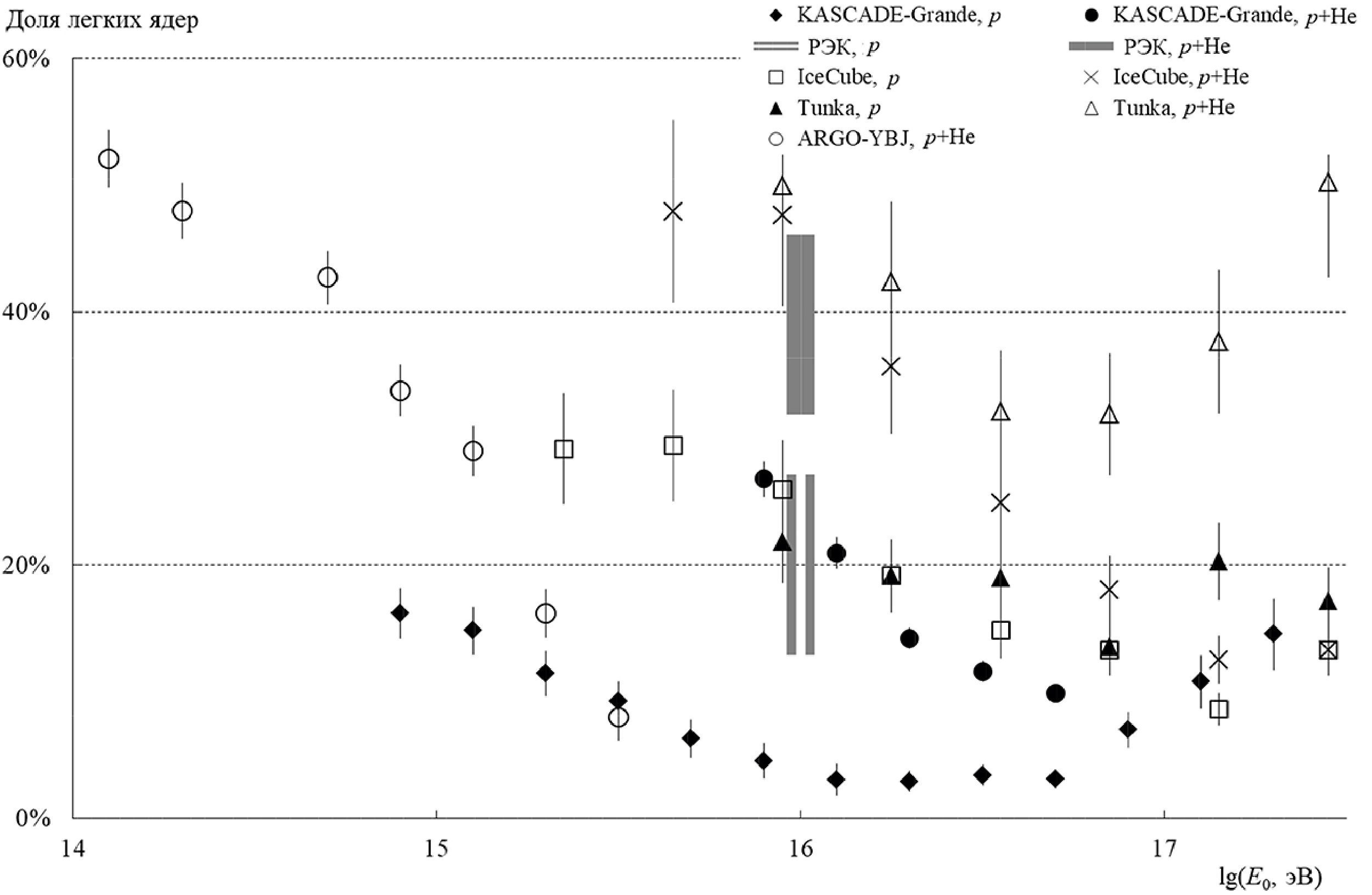}
\caption{Доля легких ядер в массовом составе ПКИ по данным экспериментов \mbox{KASCADE-Grande}~\cite{12,13}, \mbox{ARGO-YBJ}~\cite{14}, Tunka~\cite{15}, IceCube~\cite{16} и РЭК~\mbox{ПАМИР}~\cite{8}.}
\label{fig4}
\end{figure}

На рисунке~\ref{fig4} приведено изменение доли \textit{р}+Не в массовом составе ПКИ в диапазоне $E_0=1-100$~ПэВ по данным эксперимента \mbox{ПАМИР} и ряда других. Из рисунка~\ref{fig4} следует, что результаты эксперимента \mbox{ПАМИР} по оценке доли легких ядер при $E_0=10$~ПэВ на основе анализа событий с гало дают хорошее согласие с результатами экспериментов Tunka и IceCube. Из данных всех проанализированных экспериментов следует, что при энергиях ядер ПКИ $\sim10$~ПэВ наблюдается локальное утяжеление массового состава ПКИ.

\setcounter{table}{0}
\renewcommand\thetable{\arabic{table}}
\begin{table*}[htbp]
\caption{Зависимость среднего значения массового числа ядер ПКИ от $E_0$, полученная из анализа данных эксперимента \mbox{KASCADE-Grande}.}
  \centering
    \begin{tabular}[t]{|c|c|c|c|c|c|l|}
\hline    
lg$(E_0$,~eV)&$14,0-14,5$&$14,5-15,0$&$15,0-15,5$&$15,5-16,0$&$16,0-16,5$\\
\hline
$<A>$&3~(\textit{р}+Не)&5&7&10&13~(CNO)\\        
\hline
    \end{tabular}
\label{tabl:1}    
\end{table*}

В частности, по данным эксперимента \mbox{KASCADE-Grande} выполнен анализ массового состава ПКИ~\cite{8}, результаты которого приведены в таблице~\ref{tabl:1}. Анализ показал, что при $E_0\geq5$~ПэВ доля протонов существенно снижается. Массовый состав ПКИ около 10~ПэВ соответствует группе CNO, что подтверждено, например, исследованиями~\cite{17}.

{\bf 6.~Моделирование доли легких ядер в РЭК}\newline
Стволы ШАЛ, регистрируемые на РГП методом РЭК, образованы преимущественно легкой компонентой массового состава ПКИ, - более 95\% всех 100-ТэВ-х семейств $\gamma$-квантов образованы ядрами \textit{р}+Не. В исследованиях эксперимента \mbox{ПАМИР} установлено, что энергия семейств $\gamma$-квантов $\Sigma E_\gamma$, измеряемая экспериментально, связана с $E_0$ соотношением $E_0=k\Sigma E_\gamma$, где $k=10$ для ШАЛ, образованных первичными протонами, с ростом до 70 для ШАЛ, образованных тяжелыми ядрами. По экспериментальным значениям $\Sigma E_\gamma$ можно оценить $E_0$ ядер легкой группы ПКИ, генерирующих ШАЛ, характеристики которых изучаются.

Для оценки доли легких ядер, стволы ШАЛ от которых регистрируются в РЭК, и изменения доля \textit{р}+He с $E_0$, по модели \mbox{МС0-FANSY} получены характеристики ШАЛ на уровне наблюдения РЭК начиная с $E_0$ несколько ТэВ, что ниже порога регистрации по $E_0$ экспериментальной установки. Экспериментальные характеристики стволов ШАЛ, полученные методом РЭК, сопоставлены посредством нейронной сети (НС) с аналогичными, модельно полученными данными. Использована НС типа многослойный перцептрон с алгоритмом оптимизации методом шкалируемых сопряженных градиентов. Наборы данных выбирались случайным образом для обучающей, критериальной и контрольной выборок. Контроль качества обучения нейронной сети выполнен по проценту неверных предсказаний.

Для целей решения задач моделирования, ядра ПКИ разделены на две группы, - легкие \textit{р}+He и тяжелые. Переменная «тип ядра ПКИ» принята как зависимая. К ковариатам отнесены экспериментально измеряемые характеристики семейств $\gamma$-квантов, такие как количество $\gamma$-квантов в семействе, $cos\theta$, $\Sigma E_\gamma$ и средневзвешенный по энергиям $\gamma$-квантов радиус семейства, прямо коррелированные с типами ядер ПКИ, генерировавших данные ШАЛ. В частности, по причине более высокой диссипации энергии и меньшего пробега до взаимодействия в атмосфере при заданной $E_0$, семейства $\gamma$-квантов, образованные тяжелыми ядрами ПКИ, имеют больший средний радиус, нежели средний радиус семейств $\gamma$-квантов, образованных протонами. Например, экспериментально полученный в РЭК радиус семейств $\gamma$-квантов равен $(1,94\pm0,06)$~см, что не противоречит расчетному среднему радиусу $(2,01\pm0,03)$~см для легких ядер ПКИ~\cite{1}.

{\bf 7.~Изменение доли легких ядер в массовом составе ПКИ}\newline
Ранее отмечено, что методом РЭК изучаются характеристики стволов ШАЛ, регистрируемых на РГП, такие как $E_\gamma$ и $\Sigma E_\gamma$, измеренные по $\gamma$-квантам с установленным порогом регистрации. Значению $\Sigma E_\gamma$ соответствует $E_0$ первичного ядра, значение $E_\gamma$ определяет вероятность регистрации ШАЛ, генерированного данным типом ядра. Чем выше значение $E_\gamma$, тем с меньшей вероятностью зарегистрированный ШАЛ образован тяжелым ядром по причине более высокой диссипации энергии, относительно легких ядер. Иными словами, измеряемый массовый состав ПКИ зависит от порога регистрации РЭК. При равной нулю пороговой энергии регистрации $\gamma$-квантов, регистрируемый и первоначальный массовый состав ПКИ идентичны, и при данной $E_0$ РЭК будет равновероятно регистрировать ШАЛ, образованные как легкими, так и тяжелыми ядрами.

В методе РЭК порог регистрации $\gamma$-квантов достаточно высокий и составляет несколько ТэВ, что отсеивает в первую очередь ШАЛ, образованные тяжелыми ядрами ПКИ. При низких $E_0$ установка работает как сепаратор легкой компоненты ПКИ, где регистрируется не более 5\% тяжелых ядер. Однако вероятность регистрируемых ШАЛ, образованных тяжелыми ядрами, возрастает с $E_0$, или с $\Sigma E_\gamma$, что позволяет получить информацию о доле тяжелых ядер в массовом составе ПКИ, анализируя семейства $\gamma$-квантов с более высокими значениями $\Sigma E_\gamma$.

\begin{figure}[h]
\centering
\includegraphics[width=\linewidth]{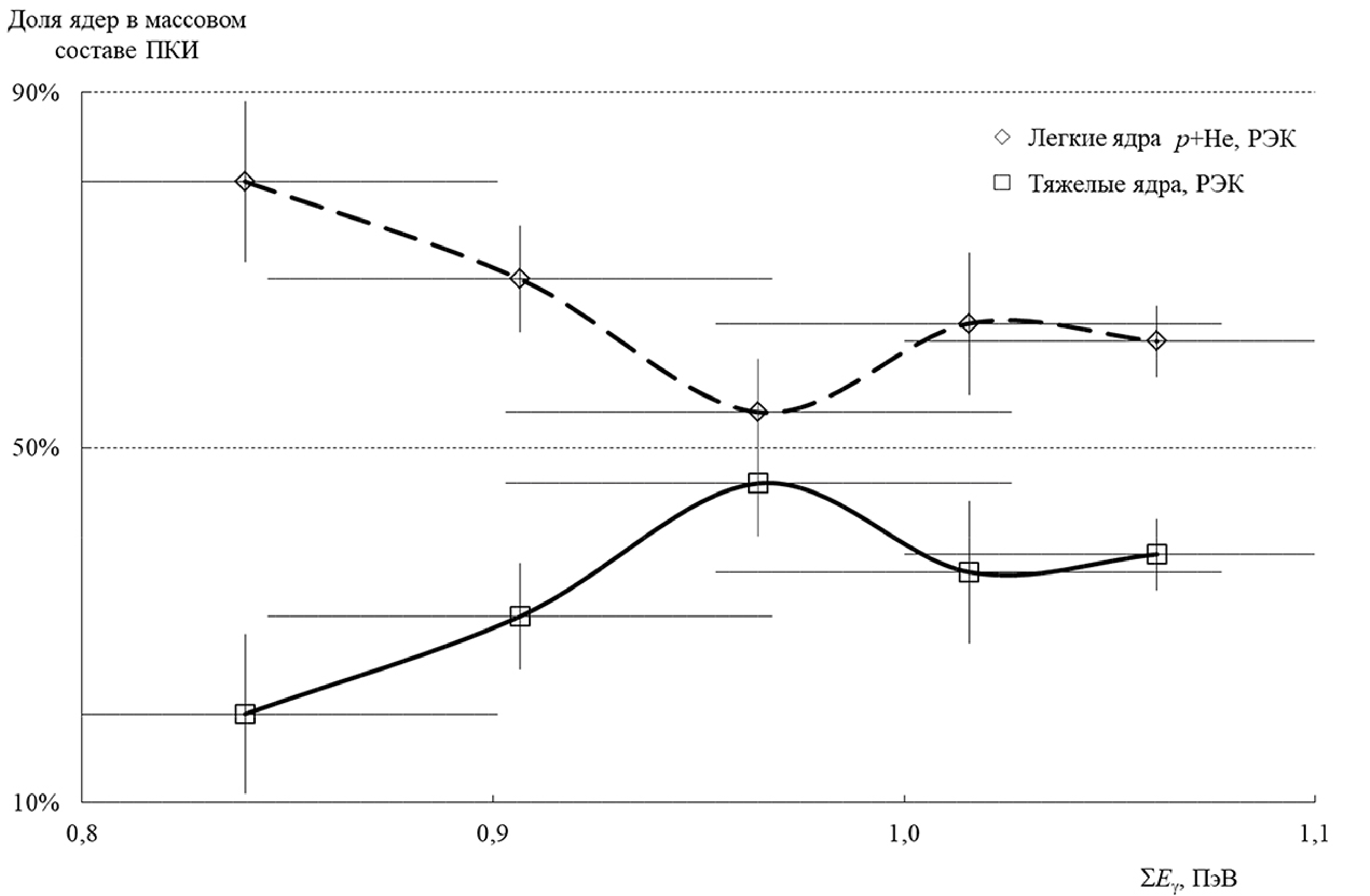}
\caption{Изменение доли легких ядер в массовом составе ПКИ с $\Sigma E_\gamma$.}
\label{fig5}
\end{figure}

Для обучения НС из базы данных (БД), содержащей модельные события, отобраны ШАЛ со значениями $\Sigma E_\gamma$ 600, 700, 800, 900 и 1000~ТэВ, полученными по пороговым $E_\gamma$, равными 4, 6, 8 и 10~ТэВ и для которых известны типы первичных ядер. Из БД экспериментальных событий аналогичным образом отобраны семейства $\gamma$-квантов для тех же значений $\Sigma E_\gamma$ и $E_\gamma$, но типы первичных ядер для которых необходимо оценить. Исследования показали, что для данной $\Sigma E_\gamma$ зависимость доли легких ядер, регистрируемой методом РЭК, от $E_\gamma$ линейна с $R_a^2>$~80\%. Из полученных регрессий оценены доли легких ядер, соответствующие нулевому порогу регистрации установки. Результаты оценок, основанных на экспериментальных данных РЭК, долей легких и тяжелых ядер в массовом составе ПКИ в зависимости от $\Sigma E_\gamma$ приведены на рисунке~\ref{fig5}.

Анализ рисунка~\ref{fig5} показывает, что доля легких ядер при низких энергиях, соответствующих 100-ТэВ-м семействам $\gamma$-квантов, та же, что и в ранее выполненных расчетах по модели \mbox{МС0-FANSY}~\cite{1} (>~90\%). Массовый состав ПКИ в рассматриваемом диапазоне энергий $\Sigma E_\gamma$ утяжеляется, однако при $\Sigma E_\gamma=800$~ТэВ наблюдается локальный минимум доли легких ядер с малой полушириной (максимум доли тяжелых ядер).

\begin{figure}[h]
\centering
\includegraphics[width=\linewidth]{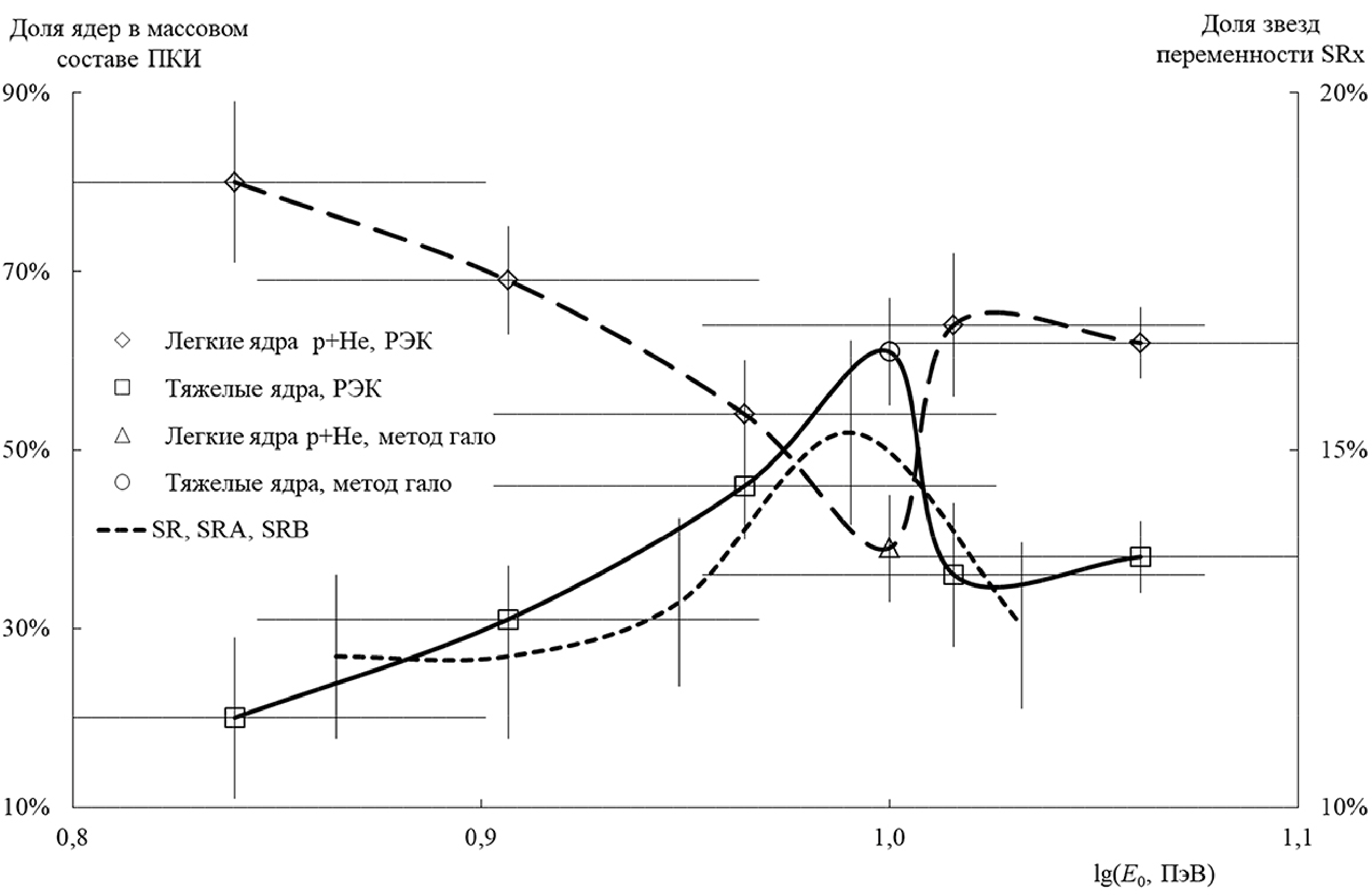}
\caption{Изменение доли легких ядер в массовом составе ПКИ (длинный пунктир) с $E_0$, в сравнении с распределением звезд переменности SRх (короткий пунктир) (процент от всех звезд, соответствующих данной $E_0$).}
\label{fig6}
\end{figure}

Переход от $\Sigma E_\gamma$ к $E_0$ выполнен по оценочным коэффициентам перехода $k$ для эксперимента \mbox{ПАМИР}. Изменение доли легких ядер в массовом составе ПКИ с $E_0$ приведено на рисунке~\ref{fig6}, где также показана минимальная доля легких ядер, полученная методом гало. Анализ рисунка~\ref{fig6} показывает, что около $E_0=10$~ПэВ наблюдается локальный максимум доли тяжелых ядер. Добавление оценки доли легких ядер методом гало уменьшает полуширину данной неоднородности. Последнее указывает на наличие локального источника ПКИ, обеспечивающего поток тяжелых ядер с энергиями $E_0=8-13$~ПэВ. Здесь также необходимо отметить, что $E_0=10$~ПэВ, - наивероятная энергия образования семейств $\gamma$-квантов с гало, вероятности образования которых приведены на рисунке~\ref{fig3}.

В исследованиях~\cite{18} показано, что нерегулярности в спектре ПКИ по $E_0$ можно объяснить распределением звезд затменно-переменного типа. В частности, спектр ПКИ по $E_0$ около 10~ПэВ сформирован звездами переменности SRx, - полурегулярными красными гигантами и сверхгигантами промежуточных или поздних спектральных классов. На рисунке~\ref{fig6} распределение звезд переменности SRx показано коротким пунктиром. Можно предположить, что звезды данного типа, - источники тяжелых ядер в массовом составе ПКИ. В частности, в исследованиях~\cite{17} показано, что источники ядер ПКИ могут находится в Галактике и обеспечивать поток ПКИ вплоть до $E_0=10^{20}$~эВ, несмотря на большой гирорадиус ядер.

Основные переменные звезды, расположенные в диапазоне по $E_0=8-13$~ПэВ, - мириды и SRx. Однако как было показано в~\cite{18}, мириды являются основными источниками ПКИ, вносящими вклад в бамп (максимум) около $E_0=100$~ПэВ, в то время как локальная нерегулярность в массовом составе ПКИ около $E_0=10$~ПэВ обеспечена звездами SRx-типа, максимум распределения которых приходится на данную $E_0$. Также помимо звезд указанных типов, в данном диапазоне по $E_0$ находится локальный источник, - звезда WR~136 спектрального класса WC5 из созвездия Лебедя. Как показано в~\cite{18}, период звезды 136-139 дней обеспечивает энергию ядер ПКИ lg($E_0$,~ПэВ)~$=0,94-0,96$, что не противоречит данным, представленным на рисунке~\ref{fig6}. Особенности звезд типа WR, - это звезды на поздних этапах эволюции, содержащие мало Н и с сильным звездным ветром. Данные звезды относительно редки, что делает звезды WR мощными и локальными по $E_0$ источниками тяжелых ядер в массовый состав ПКИ. Звезда WR~136 относится к звезде углеродной последовательности, что также не противоречит оценке массового состава около $E_0=10$~ПэВ, приведенного в таблице~\ref{tabl:1}.

{\bf 8.~Обсуждение}\newline
Тонкая структура нерегулярностей массового состава ПКИ с изменением $E_0$ в настоящее время не изучена. По данным экспериментов с высокими флуктуациями изучаемых параметров ШАЛ это сделать практически невозможно, и в данном контексте метод РЭК является уникальным. С другой стороны, необходимы специализированные программные комплексы, позволяющие создавать цифровой образ ШАЛ с низкими порогами по отслеживаемым частицам, с одной стороны, и высокими по $E_0$, с другой. Пример такого программного комплекса, - \mbox{МС0-FANSY}, включая моделирование прохождения ШАЛ через РЭК.

Изучение ПКИ в двух сопоставленных аспектах, - ядерно-физическом и астрофизическом, - необходимо для понимания природы и механизмов ускорения ПКИ. Исследование локального изменения массового состава позволяет сопоставить данную нерегулярность с $E_0$ и, в конечном итоге, сделать предположение, звезды каких типов переменностей их формируют. В частности, последние исследования показали, что спектр ПКИ по $E_0$ вплоть до сверхвысоких энергий может формироваться исключительно источниками, находящимися в Галактике.

{\bf 9.~Выводы}\keepwithnext
\begin{enumerate}
\item Экспериментальные данные, полученные методом РЭК, позволяют изучать характеристики стволов ШАЛ, где наблюдаются их минимальные флуктуации. Последнее позволяет исследовать тонкую структуру изменения массового состава ПКИ с $E_0$.

\item В спектре массового состава ПКИ по $E_0$ наблюдаются локальные неоднородности, образованные определенными типами звезд-источников. По данным эксперимента с РЭК, около энергии $E_0=10$ ПэВ наблюдается локальный максимум доли тяжелых ядер.

\item Источниками ядер ПКИ при $E_0$ около 10~ПэВ, сформировавшими максимум в изменении доли тяжелых ядер, являются звезды переменностей SR и WR. Вклад звезд других типов переменностей, в частности мирид, также существенен, однако мириды не формируют около $E_0=10$~ПэВ локальную неоднородность в массовом составе ПКИ.
\end{enumerate}

\end{document}